# Challenges for conflict mitigation in O-RAN's RAN Intelligent Controllers

Cezary Adamczyk

*Abstract*—The O-RAN architecture enables a more flexible and dynamic radio access network (RAN) control by separating hardware and software components. However, the distributed nature of the O-RAN architecture also presents several challenges for mitigating network control conflicts that can arise between different network elements. In this article, we identify key challenges for conflict mitigation in O-RAN networks, including reliable conflict detection, efficient maintenance of conflict mitigation configuration, optimal conflict resolution logic, testing and evaluation methodologies, and limited observability of O-RAN components. We propose solutions to these challenges, including pre-deployment conflict mitigation, conflict detection and resolution, and supervision and adaptation. The article concludes by highlighting the need for ongoing research to address these challenges and ensure effective conflict mitigation in O-RAN deployments.

## I. Introduction

The open radio access network (O-RAN) concept is an innovative approach to cellular network design that enables intelligent RAN control through the use of Near-Real-Time (Near-RT) and Non-Real-Time (Non-RT) RAN Intelligent Controllers (RICs) [1]. The O-RAN architecture is a key enabler for the next generation of wireless networks, providing increased flexibility and scalability through the use of virtualized radio access network components. xApps and rApps in Near-RT and Non-RT RICs are critical components of the O-RAN architecture, expanding the intelligent RAN control capabilities of O-RAN and enabling the rapid development and deployment of new network applications.

While O-RAN provides many benefits, it also introduces many new challenges related to conflicts between multiple agents providing RAN control. The majority of these conflicts should be mitigated a priori, i.e., avoided completely with proper configuration of the RAN control agents. However, as xApps and rApps can be developed and deployed in Near-RT and Non-RT RICs in an agile manner, new types of unexpected conflicts may emerge upon large-scale deployment. The RAN control conflicts occurring during network operation may lead to deterioration in network performance [2]. Thus, conflict mitigation in O-RAN is a critical aspect of ensuring reliable and efficient operation of future RAN.

To address the challenges of conflict detection and resolution in O-RAN deployments, it is essential to develop robust and flexible conflict mitigation mechanisms that can adapt to new types of applications and network control activities. In addition to the technical implementation aimed to mitigate conflicts, conflict management procedures need to be defined and introduced in the network management workflows to ensure the network is designed, built, and operated with conflict-less and intelligent RAN control in mind.

The subject of conflict mitigation in RAN control is extensively explored in the context of Self-Organized Networks (SON), where network steering decisions are done locally based on the observations of a RAN node. Moysen et al. [3] present a self-coordination framework, which utilizes machine learning (ML) to predict network performance and resolve conflicts between two SON functions, namely mobility load balancing (MLB) and mobility robustness optimization (MRO). This approach is not directly applicable to O-RAN since the RAN control in O-RAN is performed by RICs, which control multiple RAN nodes. As dedicated solution for O-RAN, Zhang et. al [4] proposed a team-learning scheme that employs ML to coordinate xApps. Nevertheless, the approach shown in this work requires the xApps to be developed with support of team learning, hence it cannot be applied generally to all O-RAN applications. Adamczyk and Kliks [5] defined a framework for conflict resolution in O-RAN, which is generally approach and demonstrates improvement in certain network performance indicators at the expense of a slight decrease in others. However, it does not address the issue of conflict mitigation comprehensively.

In this paper, we identify and discuss the key challenges that must be addressed to mitigate conflicts effectively in O-RAN deployments and ensure that RICs can dynamically adapt network configuration to current goals set by mobile network operators (MNOs). We propose solutions for effective conflict mitigation in O-RAN networks, which are categorized as pre-deployment conflict mitigation, conflict detection and resolution, and supervision and adaptation.

## II. Key challenges in the mitigation of RAN control conflicts in O-RAN

### A. Reliability of conflict detection mechanisms

**Challenge description:**

The complexity of the O-RAN radio control plane makes it challenging to ensure reliable conflict detection mechanisms, particularly in the context of inter-Near-RT-RIC conflicts and conflicts between Near-RT and Non-RT control loops. As Near-RT and Non-RT RICs utilize xApps and rApps to facilitate intelligent RAN control [6, 7], these components are not inherently aware of each other's decisions, leading to potential conflicts between multiple agents providing RAN control. The lack of coordination and standardization between the various components responsible for RAN control decisions can result in significant complexity within the O-RAN radio control, making it difficult to manage and mitigate conflicts effectively.

Intra-Near-RT and Intra-Non-RT RIC conflicts can occur when multiple xApps or rApps operate simultaneously, potentially interfering with each other or competing for radio resources. For example, one xApp may attempt to optimize radio resources for a particular service, while another xApp may seek to reduce power consumption. If not adequately managed, these conflicting goals can lead to suboptimal performance or even service disruptions.

Inter-Near-RT RIC conflicts can also arise when multiple RICs control different segments of the RAN, leading to conflicts due to overlapping coverage areas or interference between adjacent segments. Detecting such conflicts is challenging, requiring coordination between RICs and real-time data sharing.

In addition, conflicts can arise between Near-RT RICs and Non-RT RICs, with xApps within a Near-RT RIC potentially contradicting a policy or control decision from the rApps of the Non-RT RIC.

Regardless of the conflicting agents, reliable conflict detection is critical for conflict mitigation methods to take action and resolve the conflicts.

**Guidelines for addressing the challenge:**

Addressing the challenge of reliable conflict detection mechanisms in O-RAN requires the development of advanced algorithms and analytics that can effectively monitor and detect conflicts in real-time, as well as the deployment of robust data sharing and communication mechanisms between different xApps and rApps. This may involve the use of ML and artificial intelligence (AI) techniques to identify and mitigate conflicts automatically, as well as the development of standardized interfaces and protocols for exchanging network state information between different RICs.

*B. Maintenance of the conflict management configuration in an evolving network*

**Challenge description:**

The challenge of maintaining the conflict management configuration in an evolving O-RAN network arises due to the need to update the configuration with the evolution of the network and the addition or modification of xApps and rApps in the environment. This maintenance is necessary to ensure the continued reliable and efficient operation of the network. However, manual updates of the conflict management configuration can be time-consuming and error-prone, especially in large and complex networks. Additionally, the conflict management configuration must consider not only conflicts between the RAN control agents but also between the conflict mitigation mechanisms themselves – implemented CM solutions must not interfere with each other and manage the conflicts in a consistent manner. Therefore, it is essential to develop efficient and automated methods for maintaining and updating the conflict management configuration.

**Guidelines for addressing the challenge:**

To support efficient maintenance of the conflict mitigation configuration, it is necessary to invest in training and support for network operators, to ensure that they have the knowledge and skills necessary to effectively manage conflicts in O-RAN deployments. In this context, it might be necessary to adapt capabilities of operational support systems (OSS) for O-RAN network management to incorporate conflict management capabilities.

*C. Optimal conflict resolution logic*

**Challenge description:**

Ideally, the conflict resolution logic must be able to analyze the potential solutions to a conflict and select the best option based on several factors, including network performance, service availability, and user experience. Additionally, the conflict resolution logic must be able to adapt to changes in the network environment, including the addition or removal of network components, to ensure that conflicts are resolved optimally. This challenge requires the development of sophisticated algorithms and decision-making processes that can analyze complex scenarios and make informed decisions in real-time.

Finding and applying optimal conflict resolution logic can be particularly challenging in complex O-RAN radio control scenarios where there are multiple conflicting objectives, such as minimizing interference, optimizing resource allocation, and ensuring quality of service. In such scenarios, finding a solution that optimizes all objectives simultaneously may not be feasible, and trade-offs may need to be made between conflicting objectives. The chosen trade-off should be aligned with the current policies applied by the MNO.

Additionally, the optimal conflict resolution logic may vary depending on the network conditions, such as the number of active users, their location, and the type of services they are using, making it difficult to develop a single optimal logic that works in all scenarios. Finally, the real-time nature of the radio access network also adds to the complexity, as conflicts may need to be resolved quickly to avoid service disruptions, while ensuring that the resolution does not introduce new conflicts or negatively impact other network objectives.

**Guidelines for addressing the challenge:**

A promising approach to solving the issue of applying optimal conflict resolution logic in O-RAN radio control is to develop a systematic and automated method, potentially using ML and AI techniques, that can identify the root cause of the conflict and select the best resolution strategy based on a large set of factors. These could involve network parameters, traffic patterns, applied policies, and MNO's priorities.

Another approach could be to use game theory to model and analyze the interactions between the different RAN control agents and develop optimal conflict resolution strategies based on the resulting game-theoretical models. This method would require a deep understanding of the incentives and motivations of the different RAN control agents and the ability to design mechanisms that align their interests with the overall goals of the network operator.

*D. Methodologies for testing and evaluation of conflict mitigation methods*

**Challenge description:**

Defining efficient methodologies for testing and evaluating of conflict mitigation methods in O-RAN is a challenge primarily related to the lack of established standards and guidelines for testing and evaluating the performance of these methods. The complexity and heterogeneity of the O-RAN architecture make it difficult to define a uniform testing framework that can account for the unique combinations of different RAN control agents and the various interactions between them. Furthermore, the testing methodology should account for the dynamic nature of the network and the changing traffic patterns, which can affect the effectiveness of conflict mitigation methods.

In addition, the lack of objective metrics for evaluating the performance of conflict detection and resolution methods is also a major challenge. Without clear and quantitative metrics, it is difficult to compare the effectiveness of different conflict mitigation methods or to identify areas where improvements can be made. Another challenge is the lack of access to realistic testbeds that can provide a representative environment for testing and evaluating conflict mitigation methods.

**Guidelines for addressing the challenge:**

A potential way to approach this challenge would be to develop a systematic testing and evaluation framework that considers multiple scenarios of conflicting RAN control actions. This framework should take into account the dynamic nature of the network and be able to adapt to changing traffic patterns. It should also include a range of test scenarios that cover different network topologies, traffic loads, and service requirements. To ensure the reliability and reproducibility of the results, the testing methodology should be standardized and validated using a range of testbeds that provide a realistic representation of the O-RAN network. The development of such a systematic testing and evaluation framework requires close collaboration between researchers, vendors, and mobile network operators to ensure the developed methodologies align with the requirements of all relevant parties.

One possible solution to the challenge of the lack of objective metrics for O-RAN conflict mitigation is to define a set of standard metrics that can be used to evaluate the effectiveness of different conflict mitigation methods. These metrics could include factors such as the rate of false positives and false negatives in conflict detection, the time required to detect and resolve conflicts, and the impact of conflict resolution on network performance and user experience. To ensure that the metrics are meaningful and useful, they should be developed in collaboration with industry stakeholders, including mobile network operators, equipment vendors, and research organizations. Moreover, the metrics should be designed to be flexible and adaptable to new types of applications and network control activities.

*E. Limited observability of network control agents*

**Challenge description:**

The limited observability of network control agents is another significant challenge in O-RAN conflict mitigation. Network control agents are responsible for making decisions related to network configuration and optimization, but their internal operations and decision-making processes are often opaque to other agents in the system. Given the distributed nature of O-RAN architecture, it can be challenging to gather sufficient information about the network to effectively detect and resolve conflicts. This lack of visibility can make it difficult to identify conflicts and understand the root causes of those conflicts.

In addition, some network control agents may not be designed to report certain information, which limits the observability of the network. For example, a Near-RT RIC may not be designed to report specific details about its decision-making process or the state of its internal variables, which can make it difficult to determine if conflicts are arising due to a specific decision or variable. Overall, limited observability can hinder the ability to diagnose and resolve conflicts, leading to inefficient and/or ineffective conflict mitigation.

**Guidelines for addressing the challenge:**

To solve this challenge, it may be necessary to develop advanced monitoring and analytics tools that can provide deeper visibility into the behavior of O-RAN nodes and RICs, even in situations where direct observation is limited. This could involve the use of ML algorithms that can automatically detect and diagnose potential issues based on network data, or the development of specialized monitoring tools that can capture detailed information about network behavior. In addition to developing new monitoring and analytics tools, it may also be necessary to establish new protocols and standards for sharing network data and information between different O-RAN components to ensure that all network elements have the information they need to operate effectively.

III. PRINCIPLES FOR SOLVING CONFLICT MITIGATION CHALLENGES

To address all of the described challenges in O-RAN conflict mitigation, a systematic and comprehensive approach is required to be implemented widely across the telecommunications industry. This approach should take into account the various aspects of the O-RAN standard and should be based on the following principles:

1) **Optimization** – Conflict mitigation capabilities of O-RAN must not worsen and should improve, in majority of cases, network performance compared to network performance without CM measures in place. While it may not be possible for any CM solution to not worsen performance or reliability from the perspective of all metrics, the trade-off between improved and deteriorated metrics must be as beneficial to the network as possible.

2) **Standardization** – Any introduced solutions, e.g., O-RAN architecture extensions and new protocols, need to be standardized to fit in the existing O-RAN landscape. It needs to be ensured that the issue of control conflicts is approached uniformly across the industry and the CM solutions and other software O-RAN components provided by various vendors are compatible with each other.

3) **Automation** – The operation of conflict mitigation in O-RAN and its maintenance need to be as self-operating as possible. This includes evaluating and monitoring CM performance and reconfiguring CM to improve efficiency or adapt it to handle new control agents. Efficient automation in the context of O-RAN CM is likely to be enabled with ML and AI techniques, utilization of which are already envisioned in O-RAN's RICs.

4) **Objectivity** – The industry needs to develop objective metrics for evaluating the reliability of conflict detection and resolution methods and incorporate these metrics into the testing and evaluation processes that cover a wide range of O-RAN use cases and deployment scenarios.
5) **Collaboration** – The cooperation between MNOs, software vendors, and research institutions needs to be encouraged to share ideas and expertise in implementing efficient conflict mitigation solutions.

## IV. PROPOSED SOLUTIONS TO CONFLICT MITIGATION CHALLENGES

### A. Conflict mitigation activity categorization

To address the issue of RAN control conflicts in O-RAN in a practical way, we propose the following categorization of CM activities:
1) **Preventive conflict mitigation** – Activities aimed at avoiding conflicts before they occur,
2) **Conflict detection and resolution** (CD&R) – Activities aimed at detecting conflicts if they occur despite preventive measures and resolving them in an optimal way,
3) **Supervision and adaptation** (S&A) – Activities aimed at monitoring, maintaining, and reconfiguring conflict mitigation components to ensure proper network operation and alignment with policies applied by the MNO maintaining the network.

Introducing this categorization should simplify communication between parties working on issues regarding conflict mitigation.

### B. Extensions of the existing O-RAN architecture

Several new logical entities are introduced in the Near-RT RIC's Conflict Mitigation component to ensure efficient and reliable operation of the network.

Firstly, the **Conflict Mitigation** (CM) **Supervisor** is a component that monitors conflict mitigation activities and tracks their impacts on the network. Its role is to provide reports on the current conflict mitigation configuration and implement necessary updates to adapt it to current network conditions. It also adapts the configuration of other CM components to align their configuration with the current policy provided by the Non-RT RIC. As such, it is the primary agent to execute CM S&A activities.

Secondly, **Performance Monitoring** (PMon) is a component that plays a vital role in monitoring and analyzing the network's performance in real-time to identify any anomalies or deviations from expected behavior and report them to relevant CM components.

The **Conflict Detection** (CD) **Agent** is a logical component responsible for detecting conflicts between the various RAN control agents in the network. It may use different techniques, such as anomaly detection, correlation analysis, and pattern recognition, to identify conflicts between RAN control agents and raise alarms to trigger conflict resolution activities.

Finally, the **Conflict Resolution** (CR) **Agent** is responsible for resolving conflicts identified by the CD Agent. It uses a set of algorithms to determine the best course of action to resolve the conflict in an optimal way and implement the necessary configuration changes in the network.

Other than introduction of the new components, our proposal for O-RAN architecture includes a recommendation for RICs to exchange information regarding conflict mitigation activities performed within each Near-RT RIC. To accomplish this, we suggest that each Near-RT RIC shares this information via the A1 interface into the Non-RT RIC, as part of the A1 enrichment information. From there, the information can be distributed to all of the other Near-RT RICs. By incorporating this capability, the Near-RT RIC can have insight into the control decisions of other Near-RT RICs and the Non-RT RIC. This enables reliable detection and resolution of conflicts with knowledge about all relevant network elements and the conflict mitigation activities relevant to the RAN nodes managed by the Near-RT RIC.

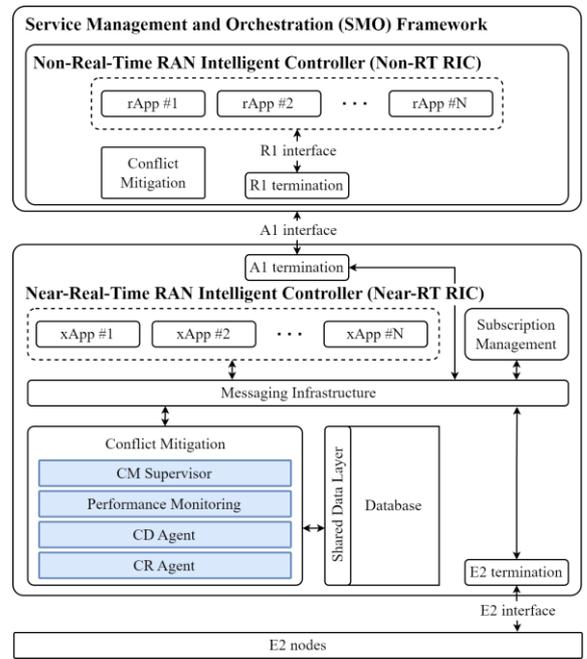

Fig. 1. O-RAN architecture with inclusion of the proposed extensions. Added components are highlighted in light blue color.

The O-RAN architecture with the proposed extensions is shown on a diagram in Fig. 1.

### C. Preventive conflict mitigation control loop

Activities that are part of the preventive conflict mitigation control loop aim to avoid RAN control conflicts before they occur in a live network. This involves both pre-deployment activities and dedicated RAN control modes that limit the possibility of conflicts.

As a key pre-deployment activity, we propose assessing impact of an application (xApp or rApp), that is being considered for deployment, in a lab or simulated environment. Ideally, the simulated network should align with the concept of digital twin, meaning that the digital representation of the network should mirror the network's behavior. This approach uses CD&R capabilities of the proposed conflict mitigation toolset to detect any conflicts happening in the network upon adding a new application to the environment. Assessing the impact of each application considered for deployment provides MNOs with insight into the pros and cons of deploying a specific application. With this knowledge, the MNO can decide to either not deploy the application, reconfigure the existing application setup in the network to remove any potentially conflicting applications, or deploy the new application regardless of any conflicts, accepting any drawbacks.

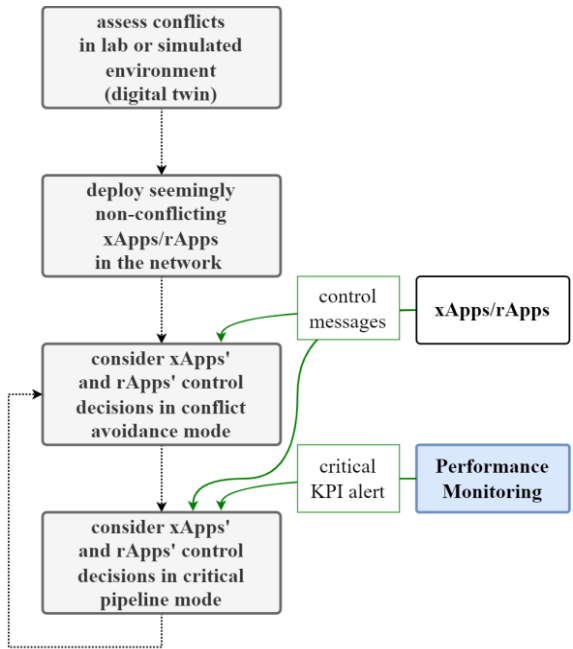

Fig. 2. Preventive conflict mitigation control loop

To avoid conflicts in a live network with multiple applications deployed, we propose two dedicated modes of operation for the CM component of RICs. The first mode is the conflict avoidance mode, in which each control message provided by an application in a RIC is first provided to the CM component. Then, the logic of the CM component (embedded within the logic of the CR Agent) decides whether the control decision should become active in the network, taking into account the network performance, prior control decisions, and the current policies set by the MNO. The conflict avoidance mode aims to limit the possibility of conflict happening in the network by ensuring that the CM component can assess and block each control decision before it is applied in the live network.

The second dedicated mode of operation of the CM components is the critical pipeline mode. Upon receiving a critical KPI alert from the Performance Monitoring component, the CM Supervisor can enforce a specific order in which xApps/rApps provide their control decisions to mitigate the issue that triggered the KPI alert. This mode of CM operation is aimed at enabling efficient network reconfiguration upon critical deterioration of the network performance. For example, the applications implementing self-healing capabilities of the network may be prioritized in the critical pipeline mode.

A diagram illustrating the preventive conflict mitigation control loop is shown in Fig. 2

### D. Conflict detection and resolution control loop

The CD&R control loop utilizes the CD Agent and CR Agent to detect and optimally resolve all types of conflict types happening in the network. The CD Agent monitors the control decisions of RAN control agents and the network itself for potential conflicts and reports them to the CR Agent. The CR Agent then analyzes the conflict reports provided by the CD Agent and decides on the best course of action to resolve the conflicts. The CR Agent may use various methods, such as prioritization, limitation, and cooldown, to resolve the conflicts. Both agents widely utilize the RIC's database to track statistical trends about conflict detection and resolution.

In the prioritization approach, the CR Agent prioritizes the control messages of the higher-ranking applications over those of lower-ranking applications to resolve the conflicts. This priority level is based on the applications' criticality in the network. The specific order of priority can be either configured statically, where it does not change unless manually modified by the MNO, or modified dynamically according to the current policy set by the MNO.

The limitation method works by limiting the capabilities of applications to modify specific parameters within a predefined range of allowed values. This approach aims to reduce the negative impact on other applications that may be affected by a particular application's actions.

Finally, in the cooldown method, one of the applications involved in the conflict is given priority over the other. Upon conflict, the control decision provided by the prioritized application is the only one that takes effect in the network. Control decisions provided by the other application for the same control target are blocked for a limited time frame, named the cooldown period. This allows the prioritized application to operate without interference from the other application and allowing its decision to have an effect on the network without disruption from other applications. After the cooldown period expires, all applications can provide control decisions again, and the conflict resolution process resumes. The duration of the cooldown period can be configured by the MNO based on the network conditions and the impact of the conflict on the network.

The control loop for CD&R activities is shown in Fig. 3.

Adamczyk and Kliks [5] implemented a conflict detection and resolution control loop, which employed the prioritization approach of the CR Agent to resolve indirect conflicts between MLB and MRO xApps. The results of their work show a mostly positive impact of CD&R activities on the network performance.

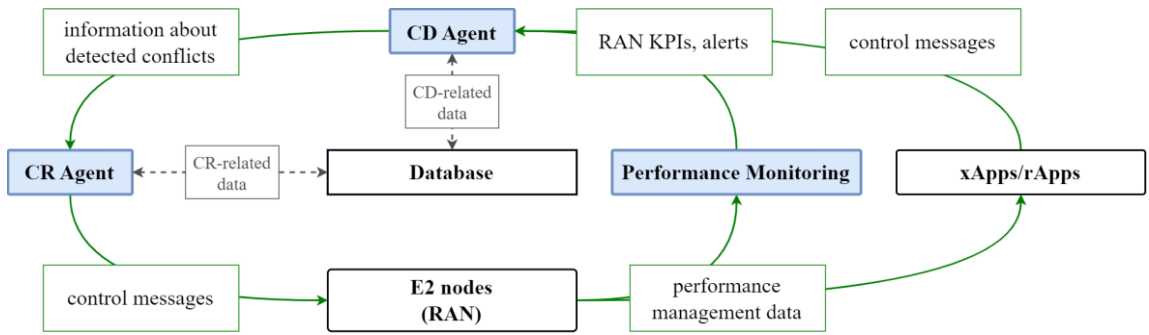

Fig. 3. Conflict detection and resolution (CD&R) control loop

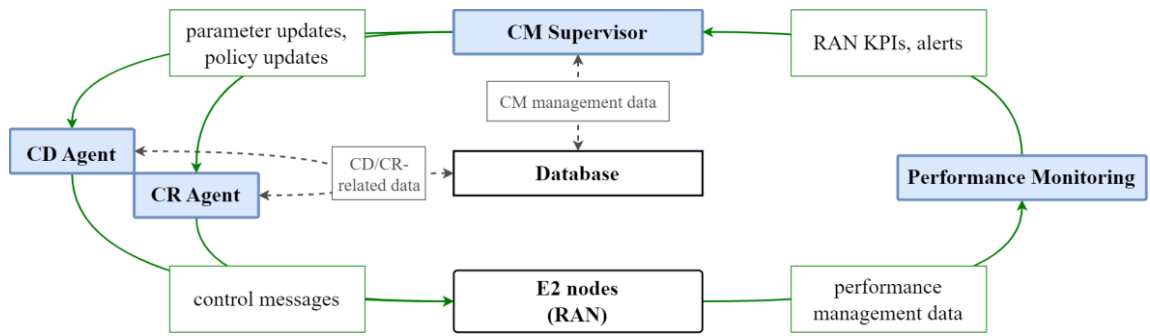

Fig. 4. Supervision and adaptation (S&A) control loop

TABLE I
MATRIX OF ALIGNMENT BETWEEN SOLUTIONS AND CHALLENGES

|  | II.A | II.B | II.C | II.D | II.E |
|---|---|---|---|---|---|
| IV.B | ✓ | ✓ | ✓ |  | ✓ |
| IV.C | ✓ | ✓ | ✓ | ✓ |  |
| IV.D | ✓ | ✓ | ✓ |  |  |
| IV.E | ✓ | ✓ | ✓ | ✓ | ✓ |

*E. Supervision and adaptation control loop*

The S&A control loop uses the CM Supervisor component to provide closed-loop monitoring and reconfiguration of the conflict mitigation component in the network. The CM Supervisor tracks the results of how the CM components solve RAN control conflicts, identifies any potential improvements, and steers how the conflicts should be resolved in the future. The CM Supervisor also adapts the CM capabilities to align with the current policy configuration set by the MNO. The S&A control loop provides a continuous feedback loop to ensure that the network remains stable and efficient, so that any conflicts that arise are detected and resolved as effectively as possible without disrupting the network operation. The CM Supervisor can also gather statistical insight for the MNO to assess whether the current application setup in the RICs is viable and provides the expected performance gains without negatively affecting reliability.

The S&A control loop envisioned as part of the proposed CM solutions is depicted in Fig. 4.

## V. CONCLUSION

To help visualize the alignment of challenges and solutions proposed in this article, a matrix has been created and is presented in Table 1. It aligns the challenges (marked with the relevant chapter indication, i.e., from II.A to II.E) and solutions (IV.B to IV.E, where solution IV.B related to introduction the recommendation for RICs to exchange information regarding conflict mitigation activities) in columns and rows, respectively. The fields in the matrix are colored to indicate whether a solution in a row addresses a challenge in the column. A green-colored field indicates that the solution in that row addresses the challenge in that column, while a grey-colored field indicates that the solution does not address the challenge. For example, in the first row (solution IV.B), the fields under challenges II.A, II.B, II.C, and II.E are colored green, indicating that this solution addresses these challenges. The field under challenge II.D is grey in the row for solution IV.B, indicating that this solution does not address this challenge. Similarly, the other solutions (IV.C, IV.D, and IV.E) are aligned in subsequent rows with their respective fields colored to indicate which challenges they address.

The proposed solutions address the challenges of conflict mitigation in O-RAN by providing preventive measures, detection and resolution mechanisms, and supervision and adaptation capabilities. Preventive CM measures help assess conflicts before deployment of new applications and provide CM modes to avoid conflicts before they occur. Conflict detection and resolution mechanisms enables to mitigate negative impacts of the conflicts that happen in a live network. Supervision and adaptation control loop introduces a concept of the CM Supervisor to provide closed-loop monitoring and reconfiguration, with possibility of adaptation of CM components to current policies.

In terms of next steps for research in the field of conflict mitigation in O-RAN, it is important to further develop and test the proposed solutions in a variety of scenarios to ensure their effectiveness and reliability. Additionally, there is a need to investigate the trade-offs between the various proposed solutions, particularly in terms of their reliability and performance impacts on the network. Finally, the continued evolution of O-RAN will require further research into the area of CM to ensure that the network remains reliable, efficient, and secure.